\documentclass[conference]{IEEEtran}
\IEEEoverridecommandlockouts

\usepackage{cite}
\usepackage{amsmath,amssymb,amsfonts}
\usepackage{graphicx}
\usepackage{textcomp}
\usepackage{xcolor}
\usepackage{booktabs}
\usepackage{array}
\usepackage{url}
\usepackage{algorithm}
\usepackage{algpseudocode}

\def\BibTeX{{\rm B\kern-.05em{\sc i\kern-.025em b}\text{T}\kern-.1667em\lower.7ex\hbox{E}\kern-.125emX}}

\begin{document}

\title{FedQML-Edge: Compact Quantum Feature Sketches for Communication-Constrained Roadside Federated Learning}

\author{\IEEEauthorblockN{Talha Azfar}
\IEEEauthorblockA{\textit{Department of Civil and Environmental Engineering} \\
\textit{Rensselaer Polytechnic Institute}\\
Troy, NY, USA \\
azfart@rpi.edu}
\and
\IEEEauthorblockN{Ruimin Ke}
\IEEEauthorblockA{\textit{Department of Civil and Environmental Engineering} \\
\textit{Rensselaer Polytechnic Institute}\\
Troy, NY, USA \\
ker@rpi.edu}
}


\maketitle
\thispagestyle{plain}
\pagestyle{plain}

\begin{abstract}
Roadside units (RSUs) supporting connected and autonomous vehicle corridors need compact models to decide when cooperative maneuvers should be rewarded, deferred, or disabled. Raw sensor streams and neural network weight checkpoints are poorly suited to bandwidth-limited, privacy-sensitive roadside learning. This paper presents \texttt{FedQML-Edge}, a federated quantum feature-sketching pipeline for traffic-stability gating. Each RSU constructs a traffic-state summary and sends circuit inputs to a quantum computer; Pauli expectations form a nonlinear sketch processed by a logistic classifier. Only classifier updates are shared with an aggregator, whose head supports reward gating. Raw observations, vehicle records, event traces, and quantum sketches remain private. We evaluate the method using NGSIM trajectories, SUMO predictive gating with sensing noise, and IBM Quantum hardware. On NGSIM, the Pauli sketch reduces test log loss by 14.4\% relative to the strongest matched classical sketch. On SUMO, it approaches larger MLPs in stable-window recall while using 7--28 times less communication per round.
\end{abstract}


\section{Introduction}

Cooperative connected autonomous vehicles (CAV) applications depend on fast local decisions made near the road, a requirement emphasized in connected and automated vehicle architectures~\cite{shladover2018connected}. Close-gap platooning, for example, can reduce aerodynamic drag and energy consumption when traffic is smooth~\cite{lammert2014effect,al2010experimental}, but the same maneuver becomes inefficient or risk-sensitive under stop-and-go turbulence, queued intersections, double parking, or degraded spacing. A roadside unit (RSU) that observes a corridor therefore needs to answer a small operational question repeatedly: is the local flow stable enough to reward cooperative behavior, or should the system defer such incentives until conditions improve?

This task is naturally a communication-constrained roadside learning problem. Roadside sensors can support local traffic summaries at high rate, and a regional controller can aggregate updates from many RSUs~\cite{shi2016edge,mao2017survey}. However, raw sensor streams, trajectories, and event records are large and privacy-sensitive, while conventional federated learning (FL) can require each RSU to upload neural network checkpoints or gradients~\cite{mcmahan2017communication,bonawitz2019towards,kairouz2021advances}. In a CAV setting, this model-update channel shares spectrum and backhaul with coordination beacons and infrastructure telemetry, so a useful roadside learner should improve calibration without large data payloads.

These constraints motivate a small trainable model, but the relationships among traffic count, speed, variability, acceleration, density, and headway still requires a nonlinear representation. We therefore study whether a fixed nonlinear feature map can improve traffic-stability classification while keeping the federated update compact. Notably, this paper evaluates a quantum-derived feature sketch as one candidate. Each RSU constructs a six-dimensional traffic-state vector and sends only the corresponding circuit inputs to a remote quantum service. The returned 17 Pauli expectation values are processed locally by an 18-parameter logistic classifier. During federated training, each RSU uploads only the resulting 144-byte classifier update to a regional FedAvg aggregator; the 17-dimensional sketches, raw observations, vehicle-level records, and local event traces are not sent to the aggregator. The global classifier then estimates the probability of stable traffic conditions, which is used to gate downstream cooperative-maneuver rewards.

Accordingly, this paper asks whether a quantum-derived nonlinear feature sketch can improve roadside traffic-stability classification, compared with matched classical sketches and compact learned models, while keeping federated updates small across heterogeneous roadside environments.

This paper makes three contributions:
\begin{itemize}
    \item We formulate roadside traffic-stability gating as a communication-constrained FL problem with a conservative false-enable requirement, separating nonlinear representation quality from trainable update size.
    \item We construct a fixed six-qubit Pauli feature sketch with a tunable angle gain and an 18-parameter FedAvg head, yielding a 144-byte client update whose size is independent of traffic volume and observation duration.
    \item We provide a matched evaluation using public NGSIM trajectories, episode-split SUMO predictive-gating experiments with sensing noise, and 1,038 circuits executed on IBM hardware. The results show when compact quantum-derived features are competitive and when larger classical models achieve lower predictive loss at the cost of greater communication.
\end{itemize}

\section{Related Work and Motivation}

\textbf{CAV edge intelligence.}
Vehicular edge systems place computation at RSUs, base stations, and nearby edge servers to reduce latency and bandwidth relative to cloud-only architectures~\cite{shi2016edge,mao2017survey}. CAV applications are especially sensitive to these tradeoffs because sensing, prediction, and coordination workloads must react to local traffic conditions~\cite{chen2019f}. RSUs can receive traffic-state summaries from loop detectors, connected-vehicle messages, radar, lidar, cameras, or fused sensing pipelines. This paper treats that sensing layer as upstream and evaluates the learning and communication layer that consumes six traffic features.

\textbf{Federated learning at the roadside.}
FL allows clients to train a shared model without transmitting raw samples~\cite{mcmahan2017communication}. Production FL systems nevertheless face synchronization, client heterogeneity, privacy leakage through updates, and bandwidth constraints~\cite{bonawitz2019towards,kairouz2021advances}. These concerns are magnified for RSUs: nodes may observe non-identical traffic distributions, connectivity can vary by corridor, and model-update traffic competes with operational CAV messages. Recent IoV/ITS work makes a similar case for federated learning when vehicular data should remain distributed~\cite{manias2021making}. \texttt{FedQML-Edge} follows the FedAvg pattern but constrains the trainable head to 18 scalar parameters. Classical random-feature and kernel-approximation methods provide important fixed-representation baselines for this question~\cite{rahimi2007random,williams2000nystrom,drineas2005nystrom}; we therefore compare the Pauli sketch with random Fourier, polynomial, and Nystr\"om/RBF sketches of the same dimension.

\textbf{Quantum feature maps.}
Quantum methods are increasingly studied for ITS tasks such as routing, autonomous driving, and quantum machine learning~\cite{zhuang2024quantumITS}; quantum-enabled FedML for roadside traffic gating remains comparatively open. Quantum feature maps project classical data into a Hilbert space where Pauli measurements can expose nonlinear structure~\cite{schuld2019quantum,havlicek2019supervised}. At the same time, NISQ execution is sensitive to noise and two-qubit depth, motivating shallow circuits and hardware-aware designs~\cite{azfar2026shallow}. Our design therefore uses the quantum circuit as a fixed sketching transform and trains a lightweight classical head. This paper contributes a controlled evaluation of the quantum-derived representation against matched classical sketches under a roadside federated-update budget, connecting quantum feature extraction with compact FedML and CAV reward gating. This makes the method compatible with local statevector simulation, Qiskit/IBM Runtime execution~\cite{qiskit2021}, and edge deployments that favor compact fixed representations over long hardware-in-the-loop optimization cycles.

\section{System Model}

\begin{figure}[t]
\centering
\includegraphics[width=\linewidth]{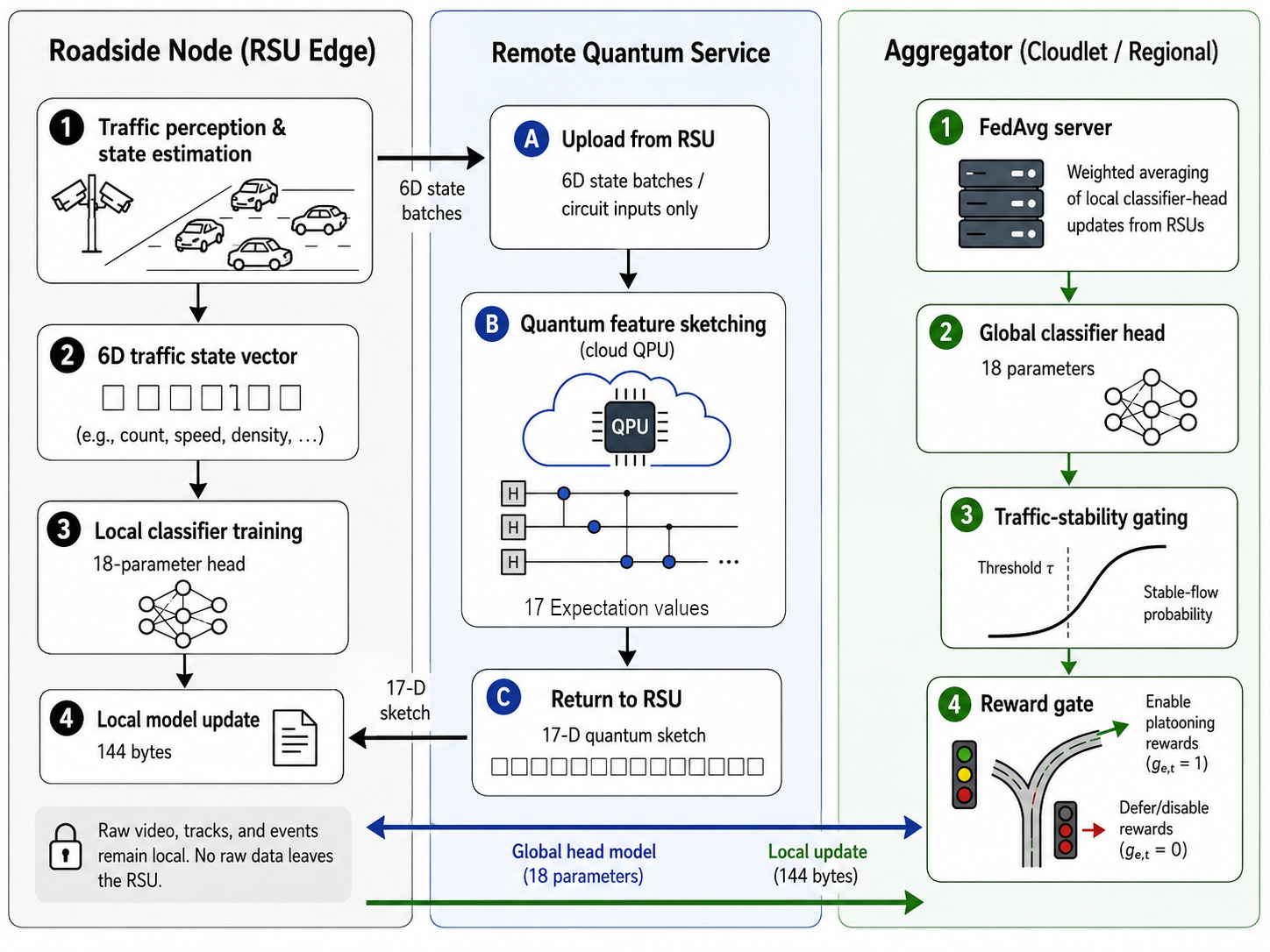}
\caption{\texttt{FedQML-Edge} computation and communication workflow. Each RSU performs traffic-state construction, local classifier training, and model-update generation. Normalized six-dimensional circuit inputs are sent to a remote quantum service, which returns a 17-dimensional Pauli
sketch to the originating RSU. The sketch is processed locally and each RSU uploads only its 144-byte classifier update, while the aggregator returns the 18-parameter global classifier head used for traffic-stability gating.}
\label{fig:architecture}
\end{figure}

Fig.~\ref{fig:architecture} summarizes the computational and communication boundaries of FedQML-Edge. Each RSU performs the classical roadside operations: traffic perception, construction and normalization of the six-dimensional traffic-state vector, local classifier training, and generation of the federated model update. Raw frames, vehicle trajectories, per-vehicle identifiers, and local event records remain at the RSU.

Quantum feature extraction is performed by remote quantum hardware. For each local traffic-state sample, the RSU transmits the six normalized circuit inputs to the quantum service. The service prepares the fixed six-qubit feature-map circuit, measures the required Pauli observables, and returns a 17-dimensional feature sketch to the originating RSU. The returned sketch is stored and processed locally, and it is not uploaded to the federated aggregator.

The RSU trains an 18-parameter logistic classifier on its local sketches and sends only the resulting 144-byte parameter vector to the regional aggregator. The aggregator performs sample-weighted FedAvg across the participating RSUs and returns the updated global classifier head. The
aggregator therefore receives neither raw traffic observations nor the 17-dimensional quantum sketches. The resulting classifier probability is used as a traffic-stability signal for downstream reward gating, while vehicle-level safety and motion control remain outside the proposed learning pipeline.

\subsection{Traffic-State Features}

At time window $t$ on edge segment $e$, the RSU constructs a six-dimensional feature vector
\begin{equation}
\mathbf{s}_{e,t} =
\left[n_{e,t}, \bar{v}_{e,t}, CV_{v,e,t}, \sigma^2_{a,e,t}, \rho_{e,t}, \bar{h}_{e,t}\right]^T ,
\end{equation}
where $n$ is vehicle count, $\bar{v}$ is mean speed, $CV_v$ is speed coefficient of variation, $\sigma_a^2$ is acceleration variance, $\rho$ is density in vehicles per 100 m, and $\bar{h}$ is mean space headway. In a deployed RSU, these values can be produced by any calibrated sensing stack, including camera-based sensing; in the public benchmark, they are computed directly from NGSIM vehicle trajectories.

The classifier estimates whether the local state belongs to a stable-traffic class that can be used as a conservative CAV maneuver-gating proxy. A deployment can mark a stable state using absolute traffic-flow thresholds:
\begin{equation}
g_{e,t}=1
\quad \text{if} \quad
CV_{v,e,t}\le\tau_v \wedge \sigma^2_{a,e,t}\le\tau_a \wedge \bar{v}_{e,t}\ge\tau_s .
\end{equation}

The gate enables platooning rewards only when the predicted stable probability exceeds a threshold $\tau$; vehicle-level safety controllers continue to enforce headway, braking, and fail-safe constraints.

\texttt{FedQML-Edge} ensures that raw observations, trajectories, and per-vehicle identifiers remain at the roadside node. The default FedAvg mode sends only the learned classifier vector. An optional statistics mode sends class counts, sketch means, and second moments for nearest-sketch inference. In both modes, update size scales with sketch dimension and remains independent of observation duration or vehicle count.

\section{Federated Quantum Sketching}

\subsection{Feature Normalization}

For reproducibility, min-max ranges are fit on the training windows and then applied to both train and test windows. In deployment, the same operation can be performed from a corridor calibration period or from local RSU ranges, element-wise:
\begin{equation}
\tilde{\mathbf{s}}_{e,t}
=
\operatorname{clip}\!\left(
\frac{\mathbf{s}_{e,t}-\mathbf{s}_{\min}}
{\mathbf{s}_{\max}-\mathbf{s}_{\min}},
0,1
\right),
\end{equation}
Degenerate feature ranges are assigned unit scale. The normalized vector directly parameterizes a six-qubit circuit.

\subsection{Quantum Feature Map}

The implemented circuit applies feature-dependent rotations and a nearest-neighbor CNOT chain. A scalar angle gain $\kappa$ controls the rotation frequency:
\begin{equation}
\begin{split}
U_{\Phi,\kappa}(\tilde{\mathbf{s}}) =
&\left(\prod_{j=1}^{5} CX_{j,j+1}\right)
\left(\prod_{j=1}^{6} R_z(\kappa\pi \tilde{s}_j)\right) \\
&\left(\prod_{j=1}^{5} CX_{j,j+1}\right)
\left(\prod_{j=1}^{6} R_y(\kappa\pi \tilde{s}_j)\right).
\end{split}
\end{equation}
The default NGSIM and hardware experiments use $\kappa=1$; the SUMO predictive-gating benchmark selects $\kappa$ on validation episodes. The output state is $|\Phi_{\kappa}(\tilde{\mathbf{s}})\rangle=U_{\Phi,\kappa}(\tilde{\mathbf{s}})|0\rangle^{\otimes 6}$. From this state, the quantum computer estimates the Pauli sketch
\begin{equation}
\mathbf{x}_{e,t} =
\left[
\langle Z_i\rangle_{i=1}^{6},
\langle Z_i Z_{i+1}\rangle_{i=1}^{5},
\langle X_i\rangle_{i=1}^{6}
\right]^T \in \mathbb{R}^{17}.
\end{equation}

The quantum service returns this 17-dimensional vector to the originating RSU. The $Z_i$ terms encode per-feature marginal structure, the adjacent $Z_iZ_{i+1}$ terms encode local feature interactions, and the $X_i$ terms add complementary-basis information. The RSU uses the sketch as the input to its local classifier. A single float64 sketch observation occupies $17 \times 8 = 136$ bytes between the RSU and the quantum service. This sketch-transfer cost is distinct from the 144-byte federated classifier update exchanged between the RSU and the
regional aggregator.

\subsection{Classifier and Federated Averaging}

The trainable model is a logistic classifier over the fixed sketch:
\begin{equation}
\hat{p}_{e,t}
=
\Pr(g_{e,t}=1\mid \mathbf{x}_{e,t})
=
\sigma\!\left(b+\mathbf{w}^{\mathsf T}\mathbf{x}_{e,t}\right),
\end{equation}
where $\mathbf{x}_{e,t}\in\mathbb{R}^{17}$ and $\boldsymbol{\theta}=[b,\mathbf{w}^{\mathsf T}]^{\mathsf T}\in\mathbb{R}^{18}$.

At round $r$, each RSU initializes from the global model
$\boldsymbol{\theta}^{(r)}$, performs $E$ local epochs, and returns
$\boldsymbol{\theta}_{e}^{(r+1)}$. The server computes
\begin{equation}
\boldsymbol{\theta}^{(r+1)}
=
\sum_{e=1}^{M}
\frac{N_e}{\sum_{j=1}^{M}N_j}
\boldsymbol{\theta}_{e}^{(r+1)},
\end{equation}
where $N_e$ is the number of samples at RSU $e$.

For FedQML and the fixed classical sketches, only the logistic-head parameters are federated; the feature map remains fixed. Raw logistic regression and MLP baselines use the same FedAvg rule on their own trainable weights. Algorithm~\ref{alg:fedqml} describes the FedQML process.

\begin{algorithm}[ht]
\caption{FedQML-Edge training}
 \label{alg:fedqml}
\begin{algorithmic}[1]
\For{each communication round $r$}
    \State Aggregator sends $\boldsymbol{\theta}^{(r)}$ to participating RSUs
    \ForAll{RSUs $e$ in parallel}
        \State Construct and normalize local traffic states
        \State Send circuit inputs to the remote quantum service
        \State Receive and locally store 17-D Pauli sketches
        \State Train the logistic head for $E$ local epochs
        \State Send $\boldsymbol{\theta}^{(r+1)}_e$ and $N_e$ to the aggregator
    \EndFor
    \State Aggregate local heads using sample-weighted FedAvg
\EndFor
\State Use the global head for traffic-stability reward gating
\end{algorithmic}
\end{algorithm}

Thus, a complete model update is 18 float64 values, or 144 bytes per RSU per round. In contrast, raw logistic regression has seven parameters and a 56-byte update, while an MLP with hidden width $H$ and one hidden layer over six inputs has $8H+1$ parameters.

For input dimension $d=6$, sketch dimension $m=17$, and MLP width of \(H\) hidden neurons, the classical feature costs are \(O(d)\) for raw and polynomial features, \(O(md)\) for random Fourier and Nyström/RBF sketches, and \(O(dH)\) per MLP forward pass. Classical statevector evaluation of the \(q=6\) quantum circuit scales as \(O(G2^q)\), while QPU execution scales approximately with circuit-gate workload and shot count, \(O(GS)\), excluding communication and queueing; therefore, this study evaluates predictive quality and update size rather than claiming a quantum runtime advantage.

\section{Experimental Setup}

\subsection{Implementation}

The prototype is implemented in Python with NumPy, Qiskit, and SUMO. Feature construction, quantum sketching, local client training, and server aggregation are implemented as separate project modules. The main NGSIM comparison uses a NumPy statevector implementation of the same six-qubit circuit for reproducibility. The SUMO benchmark runs command-line simulations over the Troy, NY network and parses 10 s \texttt{edgeData} outputs. The shot-based hardware study uses Qiskit Runtime on IBM Fez with 5000 shots per measured circuit to test whether the learned representation is robust to finite-shot and device noise.

\subsection{Benchmarks and Baselines}

We use the public FHWA Next Generation Simulation (NGSIM) vehicle trajectory dataset~\cite{fhwa2016ngsim}. We downloaded 1,000,000 trajectory rows from the official data portal and aggregated them into 10 s windows over 500 ft roadway segments. Each retained segment acts as an RSU client; windows with fewer than five unique vehicles or 30 trajectory records are discarded, and clients with fewer than eight windows are removed. This yields 519 windows across eight clients, with 363 training windows and 156 held-out test windows. The retained clients are naturally heterogeneous: the stable-window share ranges from 10.5\% on one Peachtree segment to 100\% on two short I-80 segments.

Because the downloaded NGSIM slice is heavily congested under strict absolute platooning thresholds, and because NGSIM does not provide ground-truth stability labels, we use a reproducible relative turbulence label for the benchmark. The proxy follows traffic-oscillation studies that relate unstable flow to speed fluctuations, acceleration/deceleration cycles, oscillation magnitude, and low-speed stop-and-go states~\cite{li2010measurement,zheng2011freeway,ko2006variability}. For each window, we compute
\begin{equation}
T_{e,t}=0.45R(CV_{v,e,t})+0.45R(\sigma^2_{a,e,t})+0.10(1-R(\bar{v}_{e,t})),
\end{equation}
where $R(a_{e,t})=|\{(e',t'):a_{e',t'}\le a_{e,t}\}|/N$ is the empirical percentile rank over the $N$ benchmark windows. 
The weights are benchmark choices that emphasize speed and acceleration variability while using mean speed as an indicator of congestion. The lowest 35\% of \(T_{e,t}\) values are labeled stable, yielding 177 stable and 342 unstable windows. Because this same-window label is derived from the input variables, it is used only to evaluate representation quality. The benchmark implementation also supports absolute traffic-stability thresholds for future deployment-oriented evaluations.

For the SUMO predictive-gating benchmark, we run nine episodes over the Troy, NY network~\cite{krajzewicz2012recent}. Demand scale, time interval, and random seed vary by episode. The 12 most active training edges are treated as RSU clients. Each current 10 s edge window is represented by six features derived from vehicle-count, speed, speed-change, density, and headway proxies. The target is whether the same edge remains stable over the next 20 s, using a future turbulence score based on future speed relative to the speed limit, time loss, waiting time, and density. The stable threshold is calibrated on training episodes only, and train/validation/test splits are assigned by complete episodes. A false enable occurs when the gate enables a reward for a window labeled unstable. The generated benchmark contains 2784 training windows, 1392 validation windows, and 2088 test windows; moderate feature noise is injected to approximate imperfect RSU sensing.

We compare against raw logistic regression; one-hidden-layer MLPs with widths $H\in\{2,8,16,32,64\}$; a 17-dimensional random Fourier sketch~\cite{rahimi2007random}; a 17-dimensional degree-2 polynomial sketch containing linear terms, squared terms, and adjacent interactions; and a 17-landmark Nystr\"om/RBF sketch~\cite{williams2000nystrom,drineas2005nystrom}. RBF and polynomial kernels are standard nonlinear classification baselines~\cite{cortes1995support}, and the MLP widths test whether a learned one-hidden-layer representation can obtain lower loss with more trainable parameters~\cite{hornik1989multilayer}. Each fixed sketch is followed by the same 18-parameter logistic head as FedQML. NGSIM models are trained for 100 communication rounds. SUMO raw-logistic and classical fixed-sketch models are trained for 300 rounds with five local epochs per round. SUMO FedQML selects $\kappa$ from ten angle gains between 0.5 and 3.0, $\eta$ from $\{0.2,0.4,0.7,1.0,1.5\}$, and $E$ from $\{1,2,5,10\}$ on validation episodes. The MLP grid uses learning rates 0.1, 0.3, 0.5, 0.7, and 1.0 with $E=1$ or $E=2$. Metrics include accuracy, F1 score, false-enable rate, binary cross-entropy loss, Brier score (mean squared probability error), trainable parameter count, per-round update size, and stable-window recall under a maximum false-enable constraint. We emphasize log loss and Brier score because calibrated probabilities matter for gated downstream decisions~\cite{guo2017calibration}. NGSIM gate recall is diagnostic because the threshold is selected on the held-out split; SUMO gate recall uses a validation-selected threshold frozen before test evaluation.

\section{Results}

\subsection{NGSIM Federated Benchmark}

\begin{figure}[t]
\centering
\includegraphics[width=\linewidth,trim=0 0 0 46,clip]{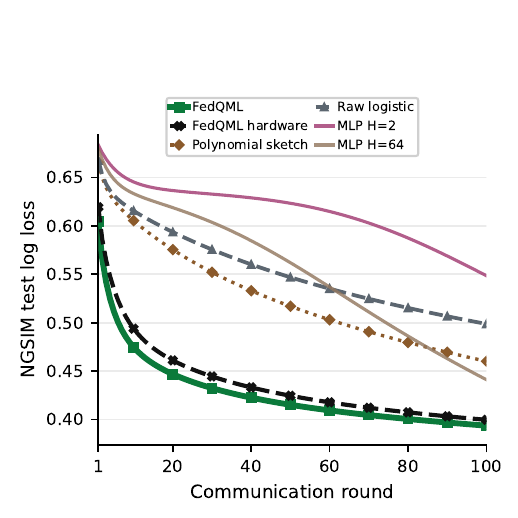}
\caption{NGSIM loss curves over 100 communication rounds. FedQML is shown for both statevector and IBM Fez hardware-derived sketches; the remaining curves show the strongest matched classical sketch, raw logistic regression, payload-matched $H=2$ MLP, and the strongest MLP.}
\label{fig:loss_curves}
\end{figure}

Fig.~\ref{fig:loss_curves} combines the NGSIM software and hardware results. \texttt{FedQML-Edge} reaches 0.394 final test log loss after 100 communication rounds, improving over matched 17-dimensional classical sketches: 0.484 for Nystr\"om/RBF, 0.460 for polynomial features, and 0.487 for random Fourier features. It also improves over raw FedAvg logistic regression (0.499), the payload-matched tiny MLP with $H=2$ (0.549), and the strongest MLP baseline, $H=64$ (0.441), while using a much smaller update. The longer 100-round setting lets the larger MLPs continue converging, but the compact Pauli sketch still gives the best NGSIM loss.

\begin{table}[t]
\centering
\caption{Fixed-Budget NGSIM Parameter, Payload, and Gate Results After 100 Communication Rounds}
\label{tab:param_payload}
\begingroup
\renewcommand{\arraystretch}{0.91}
\resizebox{\linewidth}{!}{
\begin{tabular}{lrrrrrr}
\toprule
\textbf{Architecture} & \textbf{Params} & \textbf{Update} & \textbf{Rel. Size} & \textbf{Loss} & \textbf{Acc.} & \textbf{Gate Rec.} \\
\midrule
\multicolumn{5}{l}{\textit{Fixed features}} \\
Raw logistic & 7 & 56 B & 0.39$\times$ & 0.499 & 0.763 & 38.5\% \\
\texttt{FedQML-Edge} & \textbf{18} & \textbf{144 B} & 1.00$\times$ & \textbf{0.394} & \textbf{0.827} & 42.3\% \\
Nystr\"om/RBF & 18 & 144 B & 1.00$\times$ & 0.484 & 0.731 & 21.2\% \\
Polynomial & 18 & 144 B & 1.00$\times$ & 0.460 & 0.788 & \textbf{50.0\%} \\
Random Fourier & 18 & 144 B & 1.00$\times$ & 0.487 & 0.763 & 40.4\% \\
\midrule
\multicolumn{5}{l}{\textit{Learned features}} \\
FedAvg MLP ($2$) & 17 & 136 B & 0.94$\times$ & 0.549 & 0.724 & 34.6\% \\
FedAvg MLP ($8$) & 65 & 520 B & 3.61$\times$ & 0.576 & 0.667 & 36.5\% \\
FedAvg MLP ($16$) & 129 & 1.01 KB & 7.17$\times$ & 0.504 & 0.724 & 38.5\% \\
FedAvg MLP ($32$) & 257 & 2.01 KB & 14.28$\times$ & 0.465 & 0.763 & 38.5\% \\
FedAvg MLP ($64$) & 513 & 4.01 KB & 28.50$\times$ & 0.441 & 0.776 & 44.2\% \\
\bottomrule
\end{tabular}
}
\endgroup
\end{table}

Table~\ref{tab:param_payload} reports final NGSIM results. Accuracy is measured at the default 0.5 decision threshold. ``Gate recall'' is the fraction of truly stable windows enabled after selecting the most permissive threshold that keeps false-enable decisions at or below 5\% on the held-out test split. It is a diagnostic operating-point measure for comparing classifier behavior under a conservative gate. FedQML reduces final log loss by 14.4\% relative to the strongest matched classical sketch, 21.1\% relative to raw logistic regression, and 28.2\% relative to the payload-matched $H=2$ MLP. Compared with the $H=64$ MLP, it reduces update payload by 96.5\% while improving log loss from 0.441 to 0.394. Compared with raw logistic regression, FedQML sends 88 additional bytes per update and substantially improves loss and calibration.

Fig.~\ref{fig:loss_curves} also replays the same NGSIM sketch path on IBM Fez to test finite-shot and device-noise effects. Each of the 519 windows requires two measured circuits, one for the $Z$ and $ZZ$ terms and one for the $X$ terms, giving 1038 circuits. The run used 5000 shots per circuit and 104 Runtime jobs, with at most 10 circuits per job and at most 10 active submissions. Because the sketch is fixed, the same hardware measurements support all 100 FedAvg rounds without additional quantum jobs. At round 100, hardware-derived sketches achieve 0.400 log loss and 0.846 accuracy, close to the statevector results of 0.394 and 0.827. Hardware gate recall is higher, 48.1\% versus 42.3\%, but should be interpreted cautiously because the threshold was selected on the same held-out split. Overall, the results show that current hardware can produce useful sketches for the compact FedAvg head.

\subsection{SUMO Predictive Gating Benchmark}

\begin{figure}[t]
\centering
\includegraphics[width=\linewidth]{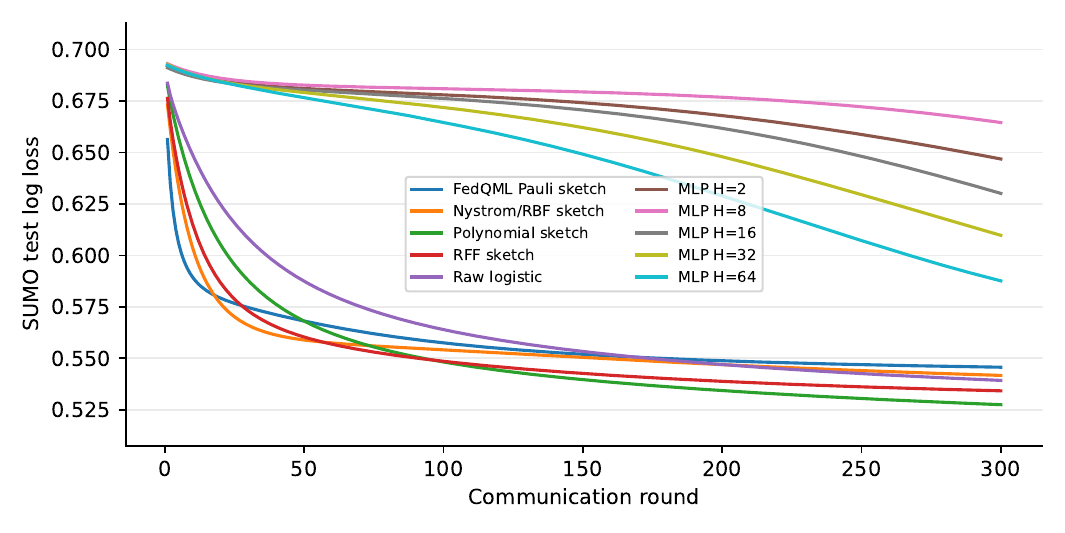}
\caption{SUMO predictive-gating test log-loss histories over 300 communication rounds. The compact fixed-sketch models improve early and then flatten, while larger MLPs continue late-round optimization with larger updates.}
\label{fig:sumo_loss_curves}
\end{figure}

Table~\ref{tab:sumo_predictive} reports the SUMO predictive-gating benchmark after 300 communication rounds, and Fig.~\ref{fig:sumo_loss_curves} shows the corresponding test-loss histories. This label is evaluated over a future horizon, making SUMO a stricter predictive test than the same-window NGSIM representation benchmark. With validation-selected $\kappa=2.5$, $\eta=1.5$, and $E=1$, \texttt{FedQML-Edge} reaches 0.534 test log loss and 30.0\% gate recall at a 4.8\% false-enable rate with a 144-byte update. It improves over raw logistic regression and matches the random Fourier sketch in loss, while polynomial features achieve lower compact-sketch loss. The operational tradeoff is strongest in gate recall: FedQML is within 0.1 percentage points of the $H=16$ MLP and 0.8 percentage points of the $H=64$ MLP while using about 7.2$\times$ and 28.5$\times$ less per-round communication, respectively.

\begin{table}[t]
\centering
\caption{SUMO predictive-gating results after 300 communication rounds. Strongest compact fixed-representation result is bold; underline marks the overall best.}
\label{tab:sumo_predictive}
\begingroup
\renewcommand{\arraystretch}{0.91}
\resizebox{\linewidth}{!}{
\begin{tabular}{lrrrrrrr}
\toprule
\textbf{Architecture} & \textbf{Params} & \textbf{Update} & \textbf{Rel. Size} & \textbf{Loss} & \textbf{Brier} & \textbf{F-E} & \textbf{Gate Rec.} \\
\midrule
\multicolumn{6}{l}{\textit{Fixed features}} \\
Raw logistic & 7 & 56 B & 0.39$\times$ & 0.539 & 0.183 & 4.5\% & 25.7\% \\
\texttt{FedQML-Edge} & 18 & 144 B & 1.00$\times$ & 0.534 & 0.180 & 4.8\% & \textbf{30.0\%} \\
Nystr\"om/RBF & 18 & 144 B & 1.00$\times$ & 0.542 & 0.184 & 4.0\% & 23.0\% \\
Polynomial & 18 & 144 B & 1.00$\times$ & \textbf{0.527} & \textbf{0.178} & 4.0\% & 28.0\% \\
Random Fourier & 18 & 144 B & 1.00$\times$ & 0.534 & 0.181 & 4.5\% & 23.7\% \\
\midrule
\multicolumn{6}{l}{\textit{Learned features}} \\
FedAvg MLP ($2$) & 17 & 136 B & 0.94$\times$ & 0.531 & 0.179 & 4.1\% & 25.7\% \\
FedAvg MLP ($8$) & 65 & 520 B & 3.61$\times$ & 0.527 & 0.179 & 4.4\% & 27.9\% \\
FedAvg MLP ($16$) & 129 & 1.01 KB & 7.17$\times$ & 0.526 & 0.178 & 4.9\% & 30.1\% \\
FedAvg MLP ($32$) & 257 & 2.01 KB & 14.28$\times$ & \underline{0.520} & \underline{0.176} & 3.6\% & 26.3\% \\
FedAvg MLP ($64$) & 513 & 4.01 KB & 28.50$\times$ & 0.521 & \underline{0.176} & 4.3\% & \underline{30.8\%} \\
\bottomrule
\end{tabular}
}
\endgroup
\end{table}

\begin{figure}[t]
\centering
\includegraphics[width=\linewidth]{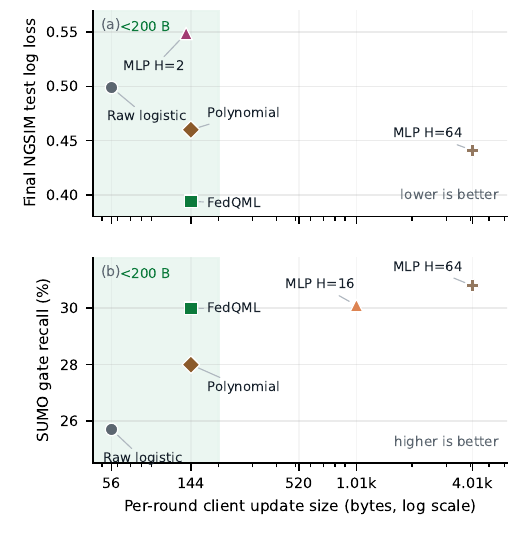}
\caption{Communication--performance tradeoffs. (a) Final NGSIM test log loss versus per-round client update size. (b) SUMO stable-window recall under the validation-selected false-enable constraint versus update size. FedQML gives the lowest NGSIM loss and approaches the SUMO gate recall of substantially larger MLPs with a 144-byte update.}
\label{fig:tradeoffs}
\end{figure}

Fig.~\ref{fig:tradeoffs} summarizes the central communication-performance message. In the NGSIM representation benchmark, FedQML gives the lowest final loss while staying in the sub-200-byte update region. In the SUMO predictive-gating benchmark, FedQML reaches nearly the same conservative gate recall as the larger MLPs while preserving the fixed-sketch payload. The FedQML angle-gain sweep is important for this result: validation-selected $\kappa=2.5$ lowers the Pauli curve without increasing update size. The tuned MLPs quantify the classical tradeoff, where lower predictive loss is available at the cost of larger client updates.

\subsection{Edge Deployment Implications}

The trainable update dimension is independent of the number of vehicles observed in the RSU window. This matters for bursty corridors because model-update load remains fixed even when observations, per-vehicle records, or queue lengths grow. The sensing layer is also separate from the federated classifier, so an RSU can improve local sensing without changing the six-feature input or sketch dimension. The current experiments use NGSIM features computed directly from public trajectories and SUMO features from microscopic edge statistics with injected sensing noise; this creates a controlled benchmark for the quantum-sketch and federated-learning layer.

\section{Discussion and Limitations}

\textbf{Why quantum sketches?}
The results position FedQML as a compact feature-sketching method evaluated through log loss, calibration, and update payload. The Pauli map offers a strong NGSIM representation and a competitive SUMO representation with a fixed 144-byte update. The SUMO comparison gives us a balanced benchmark for future quantum-edge models: matched sketches, communication-normalized curves, predictive labels, and a complete hardware sketch-extraction run.

\textbf{Client heterogeneity.}
A local-only Pauli control reaches 0.519 SUMO test loss, lower than the single global FedAvg head. This indicates strong location-specific structure among the selected SUMO clients. Accordingly, the present study evaluates communication-efficient global representation learning and motivates personalized or clustered federated extensions.

\textbf{Gate thresholding.}
For CAV maneuver gates, false-enable errors are more dangerous than false-disable errors. The threshold $\tau$ should be tuned on a validation set before deployment, and the output should inform planner rewards while vehicle-level controllers enforce motion safety. NGSIM gate recall is diagnostic because the threshold is selected on the held-out split. SUMO uses the stronger protocol: the threshold is selected on validation episodes and frozen before test evaluation. A certified controller must still enforce headway, braking, and fail-safe constraints.

\textbf{Proxy targets and splits.}
The NGSIM benchmark uses public trajectories with same-window heuristic labels, while SUMO predicts future stability using train-only calibration and episode-level splits. Together, they evaluate representation and communication tradeoffs across real trajectory data and controlled simulation. Deployment still requires held-out RSU tests, repeated scenarios, realistic sensing, and controller-grounded outcomes.

\textbf{Hardware execution.}
The IBM Fez run validates shot-based execution of the full sketch workload and produces a hardware-derived loss curve for the same FedAvg head. Because queueing, calibration drift, and provider scheduling lie outside the CAV control loop, the experiment primarily demonstrates circuit measurability and noise sensitivity. For current edge use, local simulation, quantum-inspired methods, or future dedicated accelerators are more practical.

\section{Conclusion}


This paper presented \texttt{FedQML-Edge}, a compact federated quantum feature-sketching pipeline for predictive roadside traffic-stability gating. By mapping six RSU traffic features into 17 Pauli expectation values and training an 18-parameter logistic head with FedAvg, the approach keeps raw observations local and limits each RSU update to 144 bytes. On NGSIM, FedQML achieves the lowest held-out log loss among matched FedAvg baselines. On SUMO, it improves over raw logistic regression and reaches gate recall within 0.8 percentage points of the \(H=64\) MLP while using \(28.5\times\) less communication per round. A complete 5000-shot IBM Fez run further shows that the 1038-circuit sketch workload remains useful under finite-shot and device noise, with a hardware-derived loss curve close to the statevector reference. Overall, the results position quantum-derived feature sketches as a promising compact representation for communication-constrained federated roadside learning, while motivating personalized models, held-out-RSU evaluation, and field-derived predictive labels.

\section*{Acknowledgments}
This work is funded through the IBM-RPI Future of Computing Research Collaboration. The funder played no role in study design, data collection, analysis and interpretation of data, or the writing of this manuscript. All authors reviewed the final manuscript. 

\bibliographystyle{unsrt}
\bibliography{references}

\end{document}